\documentclass[10pt,conference]{IEEEtran}

\ifCLASSINFOpdf

\else

\fi
\usepackage[colorinlistoftodos]{todonotes}
\usepackage{amssymb}
\usepackage{amsmath,epsfig,amssymb,verbatim}
\usepackage{cite}
\usepackage[caption=false]{subfig}
\usepackage{array,algorithm,algorithmic}
\usepackage{amsmath,amsfonts,amssymb}
\usepackage{algorithmic}
\usepackage{algorithm}
\usepackage{array}
\usepackage{textcomp}
\usepackage{stfloats}
\usepackage{url}
\usepackage{verbatim}
\usepackage{cite}
\usepackage[normalem]{ulem}
\usepackage{mathtools}
\usepackage{comment}
\def\BibTeX{{\rm B\kern-.05em{\sc i\kern-.025em b}\kern-.08em
    T\kern-.1667em\lower.7ex\hbox{E}\kern-.125emX}}

\usepackage{tablefootnote}
\newtheorem{theorem}{Theorem}
\newtheorem{corollary}{Corollary}

\usepackage{enumerate}

\DeclareRobustCommand{\erase}{\bgroup\markoverwith{\textcolor{red}{\rule[.5ex]{2pt}{0.5pt}}}\ULon}

\begin{document}
\bstctlcite{IEEEexample:BSTcontrol}

\title{Coexistence of OTFS Modulation With OFDM-based Communication Systems}

\author{\IEEEauthorblockN{Akram~Shafie\IEEEauthorrefmark{1},~Jinhong~Yuan\IEEEauthorrefmark{1}, Yuting Fang\IEEEauthorrefmark{2}, Paul Fitzpatrick\IEEEauthorrefmark{2},
Taka Sakurai\IEEEauthorrefmark{2}}
\IEEEauthorblockA{\IEEEauthorrefmark{1}The University of New South Wales, Sydney, NSW, 2052, Australia\\
\IEEEauthorrefmark{2}Telstra Limited, Melbourne, Australia\\
Emails:  akram.shafie@unsw.edu.au, j.yuan@unsw.edu.au,
yuting.fang@team.telstra.com,\\paul.g.fitzpatrick@team.telstra.com,~taka.sakurai@team.telstra.com}
}

\maketitle

\begin{abstract}
This study examines the coexistence of orthogonal time-frequency space (OTFS) modulation with current fourth- and fifth-generation (4G/5G) wireless communication systems that primarily use orthogonal frequency-division multiplexing (OFDM) waveforms. We first derive the input-output-relation (IOR) of  OTFS when it coexists with an OFDM system while considering the impact of unequal lengths of  the cyclic prefixes (CPs) in the OTFS signal. We show analytically that the inclusion of multiple CPs to the OTFS signal results in the effective sampled delay-Doppler (DD) domain channel response to be less sparse. We also show that the effective DD domain channel coefficients for OTFS in coexisting systems are influenced by the unequal lengths of the CPs. Subsequently, we propose an embedded pilot-aided channel estimation (CE) technique for OTFS in coexisting systems that leverages the derived IOR for accurate channel characterization. Using numerical results, we show that ignoring the impact of unequal lengths of the CPs during signal detection can degrade the bit error rate performance of OTFS in coexisting systems. We also show that the proposed CE technique  for OTFS in coexisting systems outperforms the state-of-the-art threshold-based CE technique.
\end{abstract}


\section{Introduction}

The orthogonal time-frequency space (OTFS) modulation has been envisioned as a promising candidate for realizing  reliable communications in high-mobility scenarios for fifth generation (5G) and beyond era~\cite{2021_WCM_JH_OTFS}.
The OTFS is a two-dimensional (2D) modulation scheme that carries information over the delay-Doppler (DD) domain. 
Owing to its inherent ability to handle challenges  arising from the Doppler effect, OTFS modulation has garnered considerable attention from industry and academia since its introduction in 2017~\cite{2017_WCNC_OTFS_Haddani}.

There exist two variants of OTFS, namely reduced cyclic prefix (CP)-based OTFS (RCP-OTFS) and CP-based OTFS (CP-OTFS), which differ based on the number of CPs included in the OTFS signal~\cite{2022_EmanuelBook_DDCom}. In RCP-OTFS, a single CP is included in the entire OTFS signal. Due to its ability to couple easily with the sparse DD domain representation of doubly-selective channels, RCP-OTFS has been extensively explored in the literature~\cite{2017_WCNC_OTFS_Haddani,2022_EmanuelBook_DDCom,2018_TWC_Viterbo_OTFS_InterferenceCancellation,2019_TCT_ReducedCPOTFS}.
Despite its widespread research interest, the practical implementation of RCP-OTFS faces significant challenges, mainly attributed to the utilization of rectangular pulses at the transmitter and receiver. Particularly, the use of rectangular pulses at the transmitter can cause considerable out-of-band emission (OOBE)~\cite{2022_TWC_JH_ODDM}, while including a practical filter at the transmitter and/or receiver to mitigate the effects of OOBE, can result in  
substantial performance degradation~\cite{2022_ComLet_Cheng_OTFSErrorPerformance}.

Different from RCP-OTFS, in CP-OTFS, multiple CPs are included in the OTFS signal.  Due to the presence of CPs in all transmitter pulses within the CP-OTFS signal, conventional time windowing methods such as Weighted OverLap and Add (WOLA) can be employed to mitigate the impact of OOBE in CP-OTFS.
This benefit makes CP-OTFS a viable choice for the practical implementation of OTFS. Additionally, a CP-OTFS signal can be viewed as a collection of multiple orthogonal frequency-division multiplexing (OFDM) signals, but with transmitted symbol embedding occurring in the DD domain.
This allows  the coexistence of OTFS modulation with OFDM systems, thereby enabling the seamless integration of OTFS onto current fourth-generation (4G) 
 and 5G wireless communication systems that primarily use OFDM waveforms.

The CP-OTFS was first  investigated in~\cite{2018_WCL_ModemStructureforOFDMBasedOTFS}, where its DD domain channel matrix was derived.
In~\cite{2022_IEEEAccess_Viterbo_GeneralIOR}, the time delay domain input-output relation (IOR) for CP-OTFS was derived.
Despite the progress, ~\cite{2018_WCL_ModemStructureforOFDMBasedOTFS,2022_IEEEAccess_Viterbo_GeneralIOR} did not investigate the sparsity level of the effective sampled channel response of CP-OTFS in the DD domain in comparison to that of RCP-OTFS.
Also,  the focus of~\cite{2018_WCL_ModemStructureforOFDMBasedOTFS,2022_IEEEAccess_Viterbo_GeneralIOR} was not on understanding the coexistence of CP-OTFS with current OFDM-based wireless communication systems. Thus,~\cite{2018_WCL_ModemStructureforOFDMBasedOTFS,2022_IEEEAccess_Viterbo_GeneralIOR} considered all the CPs within the CP-OTFS signal to be of equal length. 

When OTFS coexists with OFDM systems, it is crucial to account for the impacts of CP and frame structures specified in 3GPP standards~\cite{3GPPStandOFDM,5GOFDMWebpage,2018_ComStandardMag_OFDMNumerology}. Specifically, according to 3GPP standards, the downlink and uplink transmissions are arranged  into \textit{time windows} that contain multiple OFDM signals.\footnote{As per 3GPP standards, \textit{radio frames} of 10 ms duration are divided into 10 \textit{subframes}. Thereafter, subframes are further divided into two equally-sized half-frames of 0.5 ms that contain
multiple OFDM signals, which we refer to as \textit{time windows}~\cite{3GPPStandOFDM,5GOFDMWebpage,2018_ComStandardMag_OFDMNumerology}.} Within each time window, the first OFDM signal is assigned a CP of longer length than the others. This mandates the investigation of CP-OTFS with CPs of unequal lengths. 

In this work, we investigate OTFS while focusing on its coexistence with current OFDM-based 4G/5G communication systems.  We first derive the IOR of CP-OTFS while considering the impact of CPs of unequal lengths. We  show analytically that the inclusion of multiple CPs to the OTFS signal leads to every single propagation path 
being perceived as multiple taps with integer Dopplers adjacent to the true Doppler value of the corresponding propagation path. As a result of this spreading, we shown that \textit{the effective  sampled DD domain channel response of CP-OTFS can be less sparse than that of RCP-OTFS}. We also show that the effective channel coefficients for OTFS is coexisting systems are influenced by the unequal lengths of the CPs.

We next propose an embedded pilot-aided channel estimation (CE) technique that mitigates the challenges that OTFS in coexisting systems experiences during CE due to  spreading occurring along the Doppler dimension. Specifically, we propose a maximum likelihood (ML)-based CE technique that estimates the channel coefficients and Dopplers of each propagation path, using the observed channel coefficients of the multiple taps that correspond to each propagation path. 

Finally, using numerical results we show that ignoring the impact of unequal lengths of the CPs during signal detection can degrade the bit error rate (BER) 
of OTFS in coexisting systems, particularly when the CP length of the first OFDM signal within each time window is relatively much longer than the others.  We also show that the proposed CE technique for OTFS in coexisting systems outperforms the state-of-the-art threshold-based CE technique~\cite{2019_TVT_YiHong_ChannelEstimationforOTFS}, and the BER of the proposed CE technique approaches that achieved with perfect channel state information~(CSI).

\begin{figure*}[t]
\centering
\includegraphics[width=2\columnwidth]{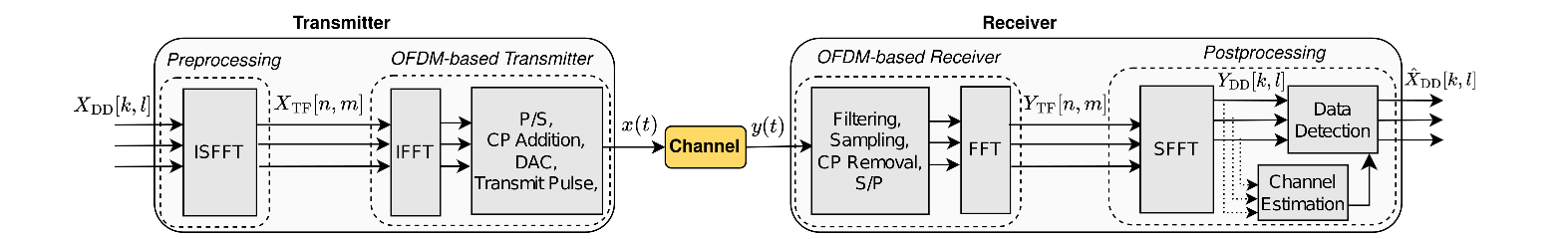}
\caption{Illustration of the system model considered in this work, where OTFS modulation coexists with current OFDM-based wireless communication systems.}\label{Fig:SysModel} \vspace{-1mm}
\end{figure*}

\section{System model}

The block diagram representation of the considered system is given in Fig.~\ref{Fig:SysModel}, where  OTFS modulation coexists with  an OFDM-based wireless communication system while introducing preprocessing at the transmitter and postprocessing at the receiver.

\subsection{Transmitter}

At the preprocessing stage of the OTFS transmitter, the DD domain symbols $X_{\textrm{DD}}[k,l]$, are converted to time-frequency (TF) domain symbols $X_{\textrm{TF}}[n,m]$, by applying inverse simplectic finite Fourier transform (ISFFT)~\cite{2017_WCNC_OTFS_Haddani}
\vspace{-1mm}
\begin{align}\label{isfft}
X_{\textrm{TF}}[n,m] = \frac{1}{\sqrt{MN}}\sum_{k=0}^{N-1} \sum_{l=0}^{M-1} X_{\textrm{DD}}[k,l]e^{j2\pi(\frac{nk}{N}-\frac{ml}{M})},
\end{align}
where \(k\!\!\in\!\!{\{0,1,..., N-1\}}\) and \(l\!\!\in\!\!{\{0,1,..., M-1\}}\) denote the Doppler and delay indices, respectively, and \(n\!\!\in\!\!{\{0,1,...,N-1\}}\) and \(m\!\!\in\!\!{\{0,1,...,M-1\}}\) denote the time and frequency indices, respectively.
Here, $M$ and $N$ denote the number of delay and Doppler bins, respectively,  in the  DD domain grid.

At the OFDM-based transmitter, TF domain symbols are used to generate $N$ OFDM signals, each having \(M\) subcarriers. Specifically,  the $n$th OFDM signal is generated as
\begin{align}\label{st2}
x_n(t)&=\sum _{m=0}^{M-1}X_{\textrm{TF}}[n,m]e^{j2\pi m \Delta {f} t} g_n(t),
\end{align}
where \(\Delta f\) denotes the subcarrier spacing, $g_n(t)$ denotes the transmitter window of the $n$th OFDM signal, given by $g_n(t)=1$, if $-T^{\mathrm{cp}}_n\leqslant t\leqslant T$, and $0$, elsewhere,
$T=\frac{1}{\Delta f}$, and $T^{\mathrm{cp}}_n$ denotes the CP length of the $n$th OFDM signal.
Finally, the time domain  OTFS signal is obtained as the delayed addition of $N$ OFDM signals, which is given by
\begin{align}\label{st}
x(t)&=\sum _{n=0}^{N-1}x_n\left(t-\left(\sum\limits_{\kappa=0}^{n}\tilde{T}_{\kappa}-\tilde{T}_{0}\right)\right)\notag\\
&=\!\!\sum _{n=0}^{N-1}\sum _{m=0}^{M-1}\!\!X_{\textrm{TF}}[n,m]e^{\!j2\pi m \Delta {f} (t-\!\!\sum\limits_{\kappa=0}^{n}\!\!a_{\kappa}\tilde{T}_{\kappa})} \!\!g_n(t-\!\!\sum\limits_{\kappa=0}^{n}a_{\kappa}\tilde{T}_{\kappa}),
\end{align}
where 
$\tilde{T}_{n}{=}T+T^{\mathrm{cp}}_n$ denotes the duration of the $n$th OFDM signal and
$a_{\kappa}{=}1$, if $ \kappa\geqslant 1$ 
 and $a_{\kappa}{=}0$, elsewhere. 

We can observe in~\eqref{st} that CPs are included in all the OFDM signals that form the OTFS signal. This variant of OTFS in which multiple CPs are included in the OTFS signal is commonly referred in the literature as CP-OTFS~\cite{2022_IEEEAccess_Viterbo_GeneralIOR,2022_EmanuelBook_DDCom}. This is different from RCP-OTFS, the widely explored variant of OTFS in which a single CP is included in the entire OTFS signal~\cite{2019_TCT_ReducedCPOTFS}.

\subsubsection{CP-OTFS with CPs of unequal lengths}

According to the New Radio (NR) frame structure that is defined in 3GPP standards, the downlink and uplink transmissions are organized into \textit{radio frames}~\cite{3GPPStandOFDM,5GOFDMWebpage,2018_ComStandardMag_OFDMNumerology}. Then the \textit{radio frames} are divided into 10 \textit{subframes}. Thereafter, subframes are further divided into two equally-sized half-frames of 0.5 ms that contain multiple OFDM signals, which we refer to as \textit{time windows}.
Within each time window, all OFDM signals, except the first, are given the same CP length. Different from the rest, the first OFDM symbol is given the CP of longer length to ensure that the number of OFDM signals within each time window is an integer. For example, for the 5G NR numerology with $\Delta f=15~\mathrm{kHz}$, the number of OFDM signals carried by a time window $S$, is 7, CP length of the first OFDM signal in each time window $T_{\mathrm{long}}^{\mathrm{cp}}$, is $5.2 \mu \mathrm{s}$, and  CP length of regular OFDM signals  $T_{\mathrm{reg}}^{\mathrm{cp}}$, is $4.69 \mu \mathrm{s}$~\cite{3GPPStandOFDM,5GOFDMWebpage,2018_ComStandardMag_OFDMNumerology}.

It is noted that the number of OFDM signals carried by a time window is typically much lower than the number of OFDM signals within an OTFS signal, i.e., $S\ll N$, especially for numerologies with low to moderate subcarrier spacing~\cite{3GPPStandOFDM,5GOFDMWebpage,2018_ComStandardMag_OFDMNumerology}.
Due to this, multiple time windows would be necessary to transmit an OTFS signal in a coexisting system.
\textit{This necessitates the investigation of CP-OTFS with CPs of unequal lengths (CP-OTFS-w-UCP).}
To account for the unequal CP length, in this work we consider $T^{\mathrm{cp}}_{n}$ to be \vspace{-1mm}
\begin{align}\label{st2}
T^{\mathrm{cp}}_{n}&=\begin{cases}
T_{\mathrm{long}}^{\mathrm{cp}}, & ~~~~~~[n]_{S}=0,\\
T_{\mathrm{reg}}^{\mathrm{cp}}, & ~~~~~~\textrm{elsewhere},
\end{cases}
\end{align}
where $T_{\mathrm{long}}^{\mathrm{cp}}> T_{\mathrm{reg}}^{\mathrm{cp}}$ and $[.]_S$ denotes the mod $S$ operation.

\subsection{Channel}

When the signal is passed through a doubly-selective channel, the complex valued basedband received signal becomes~\cite{2022_EmanuelBook_DDCom}
\begin{align} y(t) = \int_{\nu} {\int_{\tau} h(\nu,\tau)~e^{j2\pi \nu (t - \tau)}x(t - \tau) d\tau} d\nu + w(t), \label{Equ:RxSignal}
\end{align}
where $h(\nu,\tau)$ denotes the DD domain spreading function of the channel with $\nu$ and $\tau$ representing the Doppler and delay variables, respectively, and $w(t)\sim\mathcal{CN}(0,\sigma_n^2)$ denotes the complex  noise.
Assuming that the  channel is composed of $I$  separable propagations paths, $h(\nu,\tau)$ can be represented as \vspace{-3mm}
\begin{align}\label{Equ:ddchannel}
  h(\nu,\tau)=\sum_{i=1}^I h_i\delta(\nu-\nu_i)\delta(\tau-\tau_i),
\end{align}
where $h_i$, $\nu_i$, and $\tau_i$ denote the gain, Doppler, and delay of the $i$th propagation path, respectively. 
For simplicity, we consider that the Doppler and delay resolutions used to discretize the channel in the DD domain are sufficiently small  such that Dopplers and delays of propagation paths can be approximated to their nearest on-grid values~\cite{2017_WCNC_OTFS_Haddani,2018_TWC_Viterbo_OTFS_InterferenceCancellation,2022_TWC_JH_ODDM}.  Considering this,
we express $\nu_i$ and $\tau_i$ as $\nu_i {=}\frac{k_i}{NT}$ and  $\tau_i {=}\frac{l_i}{M\Delta f}$,
where $k_i$ and $l_i$ are integers representing the Doppler and delay indices on the DD domain grid for the $i$th propagation path, respectively.

\subsection{Receiver}

In practical OFDM receivers, the received signal is first passed through a band-pass filter, then the filtered signal is sampled in the time domain appropriately, and finally, all the transformations are performed in a discrete form. Following a similar process, the discrete TF domain received symbols $Y_{\textrm{TF}}[n,m]$, are obtained as \vspace{-2mm}
\begin{align}
\!\!\!\!\!\!\!Y_{\textrm{TF}}[n,m] 
=\sum_{s=0}^{M-1}\!\tilde{y}[nM{+}s]e^{{-}j\frac{2 \pi m s}{M}}\!\!\!,\label{Equ:Ymn2}
\end{align}
where $\tilde{y}[\ell]$ is the $\ell$th sample of the filtered received signal, which is obtained by sampling $\tilde{y}(t)$ at $t=\frac{T}{M}\ell{+}\!\!\sum\limits_{\kappa=0}^{\lfloor \ell/M\rfloor}\!\!a_{\kappa}T^{\mathrm{cp}}_{\kappa}$, where $\lfloor \cdot\rfloor$ denotes the floor operation.

At the receiver postprocessing stage,     TF domain received symbols are transformed to DD domain via simplectic finite Fourier transform (SFFT) \vspace{-1mm}
\begin{equation}\label{Equ:ylk1}
    Y_{\textrm{DD}}[k,l] = \frac{1}{\sqrt{NM}} \sum _{n=0}^{N-1} \sum _{m=0}^{M-1} Y_{\textrm{TF}}[n,m] e^{-j2\pi \left({\frac{nk}{N}-\frac{ml}{M}}\right)}.
\end{equation}
Finally, the signal detection and channel estimation are performed based on $Y_{\textrm{DD}}[k,l]$.

\section{Input-Output Relation}\label{section:3}


In this section, we characterize the IOR for CP-OTFS-w-UCP to understand the coexistence of OTFS with current OFDM-based wireless communication systems. 
When the received signal $y(t)$ is  bandlimited, the filtered signal of  $y(t)$ can be written as $\tilde{y}(t)=y(t)$. Based on this and ignoring the noise term, we obtain $\tilde{y}[\ell]$ 
as \vspace{-1mm}
\begin{align}
\tilde{y}[\ell] 
&= \sum_{i=1}^I \sum _{\bar{n}=0}^{N-1}\sum _{\bar{m}=0}^{M-1}h_i X_{\textrm{TF}}[\bar{n},\bar{m}]e^{j\frac{2\pi}{M}(\ell-l_i)\left(\bar{m} +\frac{k_i}{N}\right) } \notag\\
&~~~~\times e^{\!\!-j2\pi\bar{m} \left( \bar{n}{+}\!\!\sum\limits_{\kappa=0}^{\bar{n}}\!a_{\kappa}\psi_{\kappa}^{\mathrm{cp}}{-}\!\!\!\!\sum\limits_{\kappa=0}^{\lfloor \ell/M\rfloor}\!\!a_{\kappa}\psi_{\kappa}^{\mathrm{cp}}\right)} e^{j\frac{2\pi k_i}{N}\!\!\sum\limits_{\kappa=0}^{\lfloor \ell/M\rfloor}\!a_{\kappa}\psi_{\kappa}^{\mathrm{cp}}}
\notag\\
&~~~~~~\times g_n(\frac{T}{M}\ell{-}l_i{-}M(\bar{n}{+}\!\!\sum\limits_{\kappa=0}^{\bar{n}}\!a_{\kappa}\psi_{\kappa}^{\mathrm{cp}}{-}\!\!\!\!\sum\limits_{\kappa=0}^{\lfloor \ell/M\rfloor}\!\!a_{\kappa}\psi_{\kappa}^{\mathrm{cp}})), \label{Equ:ri2}
\end{align}
where $\psi_{n}^{\mathrm{cp}}=\frac{T_{n}^{\mathrm{cp}}}{T}$. 
Then substituting~\eqref{Equ:ri2} in~\eqref{Equ:Ymn2}, 
we obtain \vspace{-2mm}
\begin{align}
&\!\!\!\!\!\!Y_{\textrm{TF}}[n,m] 
=\sum_{s=0}^{M-1}\sum_{i=1}^I \sum _{\bar{m}=0}^{M-1}h_iX_{\textrm{TF}}[n,\bar{m}]e^{j\frac{2\pi k_i}{N}(n+\!\!\sum\limits_{\kappa=0}^{n}\!a_{\kappa}\psi_{\kappa}^{\mathrm{cp}})}\notag\\
&~~~~~~~~~~~~~~~~~~~~~~\times e^{-j\frac{2\pi}{M} \left( (m-\bar{m}-\frac{k_i}{N})s+l_i(\bar{m}+\frac{k_i}{N})\right)}.\label{Equ:Ymn3}
\end{align}
\noindent

We next substitute~\eqref{Equ:Ymn3} in~\eqref{Equ:ylk1} and simplify it using the steps followed similar to those in~\cite{2018_TWC_Viterbo_OTFS_InterferenceCancellation}. In doing so, we arrive at
\vspace{-1mm}
\begin{align}\label{Equ:ylk3}
    Y_{\textrm{DD}}[k,l] &= \!\!\sum_{i=1}^I h_i e^{j\frac{2 \pi (l-l_i)k_i}{NM}} \sum_{\bar{k}=0}^{N-1} \mathcal{\tilde{G}}(k,\bar{k},k_i) X_{\textrm{DD}}[\bar{k},[l-l_i]_M],
\end{align}
where
\begin{align}\label{Equ:GandF}
   \mathcal{\tilde{G}}(k,\bar{k},k_i)&=\frac{1}{N}\sum _{n=0}^{N-1} e^{-j\frac{2 \pi }{N}(n(k-\bar{k}-k_i)-k_i\!\!\sum\limits_{\kappa=0}^{n}a_{\kappa}\psi_{\kappa}^{\mathrm{cp}})}.
\end{align}

For the case when $N$ is an integer multiple of $S$, $\mathcal{\tilde{G}}(k,\bar{k},k_i)$ in~\eqref{Equ:GandF} can be simplified as
\vspace{-2mm}
\begin{align}\label{Equ:G2}
   &\mathcal{\tilde{G}}(k,\bar{k},k_i)=\frac{1}{N}\Big(\sum _{n=0}^{S-1} e^{-j\frac{2 \pi }{N}(n(k-\bar{k}-k_i(1+\psi^{\mathrm{reg}})))}\notag\\
   & ~~~~~~~~~+\!\!\sum _{n=S}^{2S-1} e^{-j\frac{2 \pi }{N}(n(k-\bar{k}-k_i(1+\psi^{\mathrm{reg}}))-k_i\psi^{\mathrm{ext}})}+\cdots\cdots\!\Big)\notag\\
   &=\frac{1}{N}\!\!\sum _{\beta=0}^{\frac{N}{S}-1}\! \sum _{\tilde{n}=0}^{S-1}\!\! e^{-j\frac{2 \pi }{N}((\tilde{n}+\beta S)(k-\bar{k}-k_i(1+\psi^{\mathrm{reg}}))-\beta k_i\psi^{\mathrm{ext}})},
\end{align}

~\\
\noindent
where $\psi^{\mathrm{reg}}{=}T_{\mathrm{reg}}^{\mathrm{cp}}/T$, and $\psi^{\mathrm{ext}}{=}(T_{\mathrm{long}}^{\mathrm{cp}}-T_{\mathrm{reg}}^{\mathrm{cp}})/T$.
Thereafter, we separate the terms on $\beta$ and $\tilde{n}$ in~\eqref{Equ:G2} and then further simplify it to arrive at
\begin{align}\label{Equ:G3}
   \!\!\!\!\!\!\mathcal{\tilde{G}}(k,\bar{k},k_i)&=\frac{1}{N}\sum _{\beta=0}^{\frac{N}{S}-1} e^{-j\frac{2 \pi S\beta}{N}(k-\bar{k}-k_i(1+\psi^{\mathrm{reg}}+\frac{\psi^{\mathrm{ext}}}{S}))} \notag\\
  & ~~~~~~~~~~~~~~~~~\times \sum _{\tilde{n}=0}^{S-1} e^{-j\frac{2 \pi \tilde{n} }{N}(k-\bar{k}-k_i(1+\psi^{\mathrm{reg}}))}\notag\\
  &=\begin{cases}
\delta(-k+\bar{k}),& k_i=0,\!\!\!\!\!\!\\
\frac{e^{-j2\pi (k-\bar{k}-k_i(1+\psi^{\mathrm{reg}}+\frac{\psi^{\mathrm{ext}}}{S}))}-1}{e^{-j\frac{2 \pi S}{N}(k-\bar{k}-k_i(1+\psi^{\mathrm{reg}}+\frac{\psi^{\mathrm{ext}}}{S}))}-1}\\
~~\times \frac{e^{-j\frac{2 \pi S}{N}(k-\bar{k}-k_i(1+\psi^{\mathrm{reg}}))}-1}{N e^{-j\frac{2 \pi }{N}(k-\bar{k}-k_i(1+\psi^{\mathrm{reg}}))}-N},& \textrm{elsewhere}.\!\!\!\!\!\!\!\!
\end{cases}
\end{align}
By analyzing~\eqref{Equ:G3}, it can be shown that the magnitude of $\mathcal{\tilde{G}}(k,\bar{k},k_i)$ has its peak around $\bar{k}=k-k_i$, and it decreases with an approximate slope of $\frac{\pi}{N}(k-\bar{k}-k_i)$.
Based on this, we consider only a small number of terms in the summation on $\bar{k}$ in~\eqref{Equ:ylk3}~\cite{2018_TWC_Viterbo_OTFS_InterferenceCancellation}. In doing so, 
we finally arrive at the IOR for CP-OTFS-w-UCP and present it in the following theorem.


\begin{theorem}\label{Thr:IORVCPOTFS}
The  DD domain received symbols $Y_{\textrm{DD}}[k,l]$ for CP-OTFS-w-UCP is given by
\begin{align}\label{Equ:ylkF}
    Y_{\textrm{DD}}[k,l] &\approx \sum_{i=1}^I h_i e^{j\frac{2 \pi (l-l_i)k_i}{NM}} \sum_{q=-\frac{\hat{N}}{2}}^{\frac{\hat{N}}{2}-1} \mathcal{G}(q,k_i)\notag\\
&~~~~~~~~~~~~\times  X_{\textrm{DD}}[[k-k_i+q]_N,[l-l_i]_M],
\end{align}
where $0\ <\hat{N} \leqslant N$ and the spreading function along the Doppler dimension $\mathcal{G}(q,k_i)$, is given by
\begin{align}\label{Equ:ylk6}
    \mathcal{G}(q,k_i)=\begin{cases}
\delta(q),& k_i=0,\\
\frac{e^{-j2\pi (-q-k_i(\psi^{\mathrm{reg}}+\frac{\psi^{\mathrm{ext}}}{S}))}-1}{e^{-j\frac{2 \pi S}{N}(-q-k_i(\psi^{\mathrm{reg}}+\frac{\psi^{\mathrm{ext}}}{S}))}-1}\!\!\!\!\!\!\!\!\\
~~~\times \frac{e^{-j\frac{2 \pi S}{N}(-q-k_i\psi^{\mathrm{reg}})}-1}{N e^{-j\frac{2 \pi }{N}(-q-k_i\psi^{\mathrm{reg}})}-N},& \textrm{elsewhere}.
\end{cases}
\end{align}
\noindent
\end{theorem}

\subsection{Special Case: CP-OTFS with CP of the equal lengths}

We next simply  Theorem~\ref{Thr:IORVCPOTFS} to obtain the element-wise IOR of CP-OTFS with CPs of equal lengths (CP-OTFS-w-ECP). 

\begin{corollary}\label{Cor:IORCPOTFS}
The DD domain received symbols $Y_{\textrm{DD}}[k,l]$ for CP-OTFS-w-ECP is given by
\vspace{-2mm}
\begin{align}\label{Equ:ylkSC}
    Y_{\textrm{DD}}[k,l] &\approx \sum_{i=1}^I h_i e^{j\frac{2 \pi (l-l_i)k_i}{NM}} \sum_{q=-\frac{\hat{N}}{2}}^{\frac{\hat{N}}{2}-1} \mathcal{G}_{\textrm{sc}}(q,k_i)\notag\\
&~~~~~~~~~~~~\times  X_{\textrm{DD}}[[k-k_i+q]_N,[l-l_i]_M],
\end{align}
where
\begin{align}\label{Equ:Gsc}
    \mathcal{G}_{\textrm{sc}}(q,k_i)=\begin{cases}
\delta(q),& ~~k_i=0,\\
\frac{e^{j2\pi(q_k+k_i \psi^{\mathrm{reg}})}-1}{N e^{j\frac{2 \pi }{N}(q_k+k_i \psi^{\mathrm{reg}})}-N},
& ~~\textrm{elsewhere}.
\end{cases}
\end{align}
\noindent
\end{corollary}

From Theorem~\ref{Thr:IORVCPOTFS} and Corollary~\ref{Cor:IORCPOTFS},  we can observe that, regardless of whether the CPs are of unequal or equal lengths, a propagation path with non-zero Doppler causes $\hat{N}$ transmitted symbols to contribute towards any given received symbol. In other words, every single propagation path with non-zero Doppler is perceived as multiple taps with integer Dopplers adjacent to the true Doppler value of the corresponding propagation paths. This shows the spreading that occurs along the Doppler dimension for CP-OTFS.
Due to this spreading, \textit{the effective sampled DD domain channel response for CP-OTFS can be less sparse as compared to that of RCP-OTFS}.

\begin{figure}[!t]
\centering\subfloat[\label{2a}RCP-OTFS]{ \includegraphics[width=0.46\columnwidth]{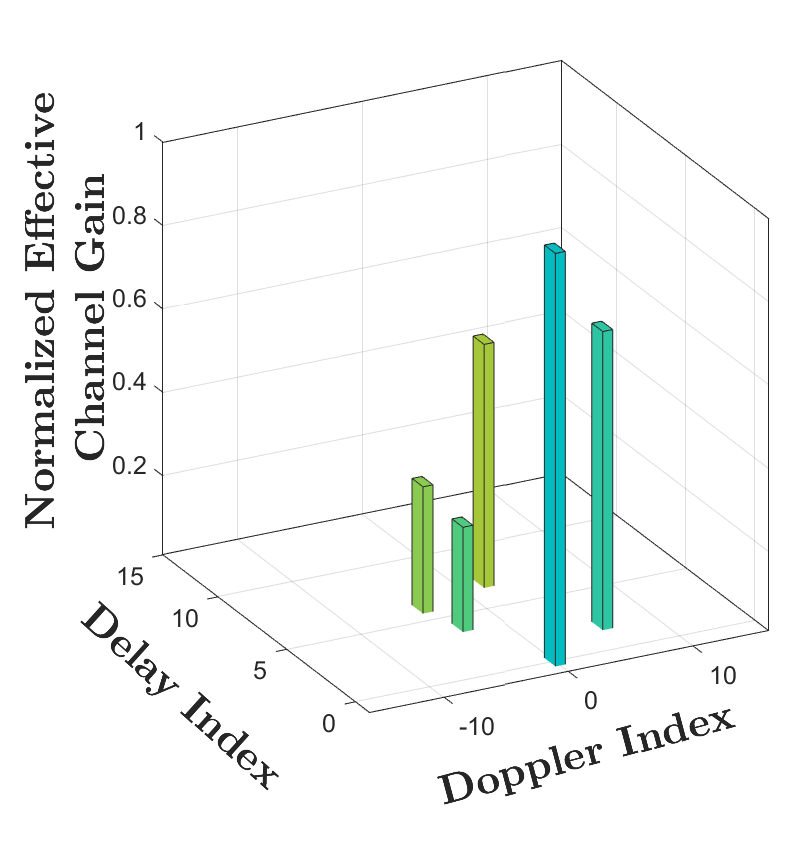}} \hspace{2 mm} \subfloat[\label{2b}CP-OTFS]{\includegraphics[width=0.46\columnwidth]{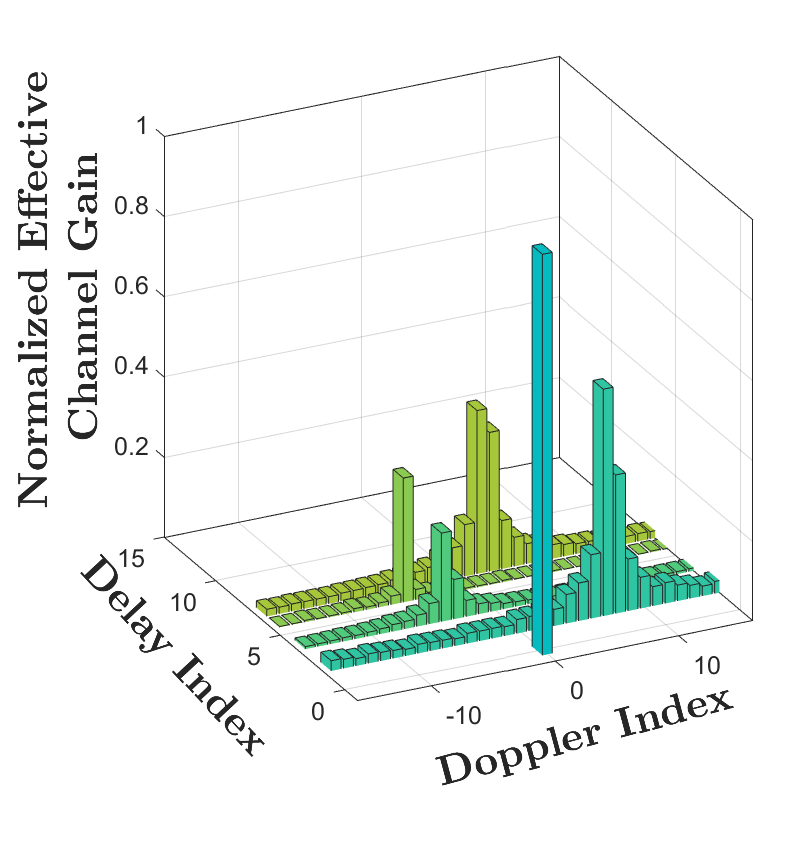}}
  \caption{Illustration of the normalized  effective sampled DD domain channel response for RCP-OTFS and CP-OTFS.}\label{Fig:ChanGainCompare}
\end{figure}

To visualize the difference between RCP-OTFS and CP-OTFS, we plot the normalized  effective sampled channel response for RCP-OTFS and CP-OTFS in Fig.~\ref{Fig:ChanGainCompare}. We consider  a doubly-selective channel that has five propagation paths, with each path characterized by on-grid delay and on-grid Doppler values, and set $N =M= 32$. 
From Fig.~\ref{Fig:ChanGainCompare}(a) we can observe that the effective sampled DD domain channel response is indeed sparse for RCP-OTFS. However, in Fig.~\ref{Fig:ChanGainCompare}(b), we can see that for CP-OTFS, channel response in the DD domain can be less sparse due to the spreading that occurs along the Doppler dimension.

By comparing Theorem 1 with Corollary 1, we note that although spreading along the Doppler dimension is experienced by both CP-OTFS-w-UCP and CP-OTFS-w-ECP, the DD domain channel coefficients of CP-OTFS-w-UCP are different from those of CP-OTFS-w-ECP. This difference is characterized in our analysis, which shows the impact of unequal lengths of the CPs on the IOR and potentially transceiver designs, such as channel estimation and detection performance.


\section{Channel Estimation}

\label{Sec:CE}

In this section, we propose an
embedded pilot-aided CE technique for OTFS in coexisting systems. 
To facilitate channel estimation, we first rearrange the OTFS frame by adding a pilot symbol, and then guard symbols surrounding the pilot symbols,  in a manner that effectively prevents interference between the pilot and data symbols in the DD domain~\cite{2019_TVT_YiHong_ChannelEstimationforOTFS}. Specifically, we let \vspace{-2mm}
\begin{align}\label{gtx}
\!\!\!\!X_{\textrm{DD}}[k,l]&\!=\!\!\begin{cases}
x_p, & \!\!k=k_p,l=l_p,\\
0, & \!\!k_p{-}2\hat{k}_{\mathrm{max}}\leqslant k\leqslant k_p{+}2\hat{k}_{\mathrm{max}},\!\!\!\!\!\!\\
 & ~~~l_p{-}l_{\mathrm{max}}\leqslant l\leqslant l_p{+}l_{\mathrm{max}},\!\!\!\\
X_{\textrm{DD,d}}[k,l], & \!\!\textrm{otherwise},
\end{cases}
\end{align}
where $x_p$ and $X_{\textrm{DD,d}}[k,l]$ denote the pilot and data symbols, respectively, and $[k_p,l_p]$ denotes the arbitrary DD domain grid location for the pilot. Also, $\hat{k}_{\mathrm{max}}=k_{\mathrm{max}}+\hat{N}$ with
$k_{\mathrm{max}}=|\nu_{\mathrm{max}}|N T$ and $l_{\mathrm{max}}=\tau_{\mathrm{max}}M \Delta f$, where $\nu_{\mathrm{max}}$ and $\tau_{\mathrm{max}}$ are the maximum of the Doppler and delay values of the propagation paths, respectively.

In the literature, threshold detection-based CE has been widely considered~\cite{2019_TVT_YiHong_ChannelEstimationforOTFS}. In this method, the received DD domain symbols which carry the pilot power, $Y_{\mathrm{ch}}[l,k]=Y_{\textrm{DD}}[l+l_p,k+k_p-\hat{k}_{\mathrm{max}}]$, where $l\in \{0,\cdots, l_{\mathrm{max}}\}$, $k\in \{0,\cdots,2\hat{k}_{\mathrm{max}}\}$, are directly used for CE. 
The threshold detection-based CE technique has been reported to be effective even with moderate pilot power when spreading does not occur along the Doppler (or/and delay) dimension, e.g., RCP-OTFS with on-grid Doppler (or/and on-grid delay).
However, in scenarios where spreading occur along the Doppler (or/and delay) dimension, 
such as that of OTFS in coexisting systems which is the focus of this work,
threshold detection-based CE technique would demand extremely high pilot power to accurately characterize the channel~\cite{2019_TVT_YiHong_ChannelEstimationforOTFS}.

To overcome this challenge, we propose an ML-based CE technique for OTFS in coexisting systems that characterizes the channel  using the IOR derived in Theorem~\ref{Thr:IORVCPOTFS} and the channel coefficients of the multiple taps that correspond to the propagation path.
In the proposed ML-based CE technique, the channel coefficient and Dopplers of propagation paths associated with delay bins $l=0, \cdots, l_{\mathrm{max}}$, are estimated one after the other. We first assume that there exists a maximum of $\bar{I}_l$ propagation paths in the $l$th delay bin.
We then estimate the channel coefficients and Doppler indices of the $\bar{I}_l$ propagation paths in the $l$th delay bin by maximizing the log-likelihood function, or minimizing the Euclidean distance as 
\begin{align}\label{Equ:L1}
   \mathcal{L}\left({\boldsymbol \theta}_{l}/\mathbf{y}_{\mathrm{ch},l}\right)&{=}\left|\left|\mathbf{y}_{\mathrm{ch},l}{-}\!\sum_{\iota=1}^{\bar{I}_{l}}h_{l,\iota} { \boldsymbol{\Psi}}_{l, \iota}x_p\right|\right|^2,
\end{align}
where $||\cdot||$ denotes the absolute operation, $\mathbf{y}_{\mathrm{ch},l}=[Y_{\mathrm{ch}}[l,0], Y_{\mathrm{ch}}[l,1], \cdots, Y_{\mathrm{ch}}[l,2\hat{k}_{\mathrm{max}}]]^{\textrm{T}}$ and ${ \boldsymbol{\Psi}}_{l, \iota}= e^{j\frac{2 \pi l_p k_{l, \iota}}{NM}}\times [\mathcal{G}(k_{l, \iota}{+}\hat{k}_{\mathrm{max}},k_{l, \iota}), \mathcal{G}(k_{l, \iota}{+}\hat{k}_{\mathrm{max}}{-}1,$ $\cdots,  $ $\mathcal{G}(k_{l, \iota}{-}\hat{k}_{\mathrm{max}},k_{l, \iota})]^{\textrm{T}}$ are  $(2\hat{k}_{\mathrm{max}}{+}1)\times 1$ vectors, $h_{l,\iota}$ and $k_{l,\iota}$ are the channel coefficient and the Doppler index of the $\iota$th propagation path in the $l$th delay bin, respectively, and ${\boldsymbol \theta}_{l}=\{h_{l, 1},h_{l, 2},\cdots,h_{l, \bar{I}_{l}},k_{l, 1},k_{l, 2},\cdots,k_{l, \bar{I}_{l}}\}$.
Mathematically, the ML estimator for the $l$th delay bin is written as 
\begin{align}\label{Equ:MLEst1}
   \hat{{\boldsymbol \theta}}_{l}&=\underset{\substack{{{\boldsymbol \theta}_{l}\in \mathbb{C}^{\bar{I}_{l}}\times \mathbb{Z}^{\bar{I}_{l}}}}}{\arg \min} \mathcal{L}\left({\boldsymbol \theta}_{l}/\mathbf{y}_{\mathrm{ch},l}\right).
\end{align}
A brute-force search in a $2\bar{I}_{l}$-dimensional
domain is necessary to find the solution to~\eqref{Equ:MLEst1}, which is unfeasible in general. In this work, we focus on the scenario where there is at most one propagation path in each delay bin and discuss the solution to~\eqref{Equ:MLEst1} accordingly.\footnote{The solution to~\eqref{Equ:MLEst1} for the scenario where there is more than one propagation path for each delay bin will be discussed in our future works.}

For the scenario where there exists at most one propagation path in each delay bin, we begin by considering $\bar{I}_{l}=1$ in~\eqref{Equ:L1}.
We then differentiate~\eqref{Equ:L1} with respect to (w.r.t.) $h_{l,1}$ and equate it to zero to obtain $h_{l,1}$  in terms of $k_{l, 1}$ as
\begin{align}\label{Equ:h1}
   h_{l,1}&=\frac{x^{\mathrm{H}}_p \boldsymbol{\Psi}^{\mathrm{H}}_{l, 1}\mathbf{y}_{\mathrm{ch},l}}{\left|\left| \boldsymbol{\Psi}_{l, 1}x_p\right|\right|^2},
\end{align}
where $(\cdot)^{\mathrm{H}}$ denotes the Hermitian transpose operation. Thereafter, we expand~\eqref{Equ:L1} while using~\eqref{Equ:h1} in it and find that the minimization in~\eqref{Equ:MLEst1} 
reduces to maximizing the function
\begin{align}\label{Equ:XX}
   &\mathcal{L}_2\left(k_{l, 1}/\mathbf{y}_{\mathrm{ch},l}\right)=\frac{\left|\left|x^{\mathrm{H}}_p {\boldsymbol{\Psi}}_{l,1}^{\mathrm{H}}\mathbf{y}_{\mathrm{ch},l}\right|\right|^2}{\left|\left|\boldsymbol{\Psi}_{l, 1}x_p\right|\right|^2}.
\end{align}
Thus, the ML estimator to identify $k_{l, 1}$ can be written as \vspace{-2mm}
\begin{align}\label{Equ:MLEst2}
   \hat{k}_{l, 1}&=\underset{\substack{{k_{l, 1}\in \mathbb{Z}}}}{\arg \max}~ \mathcal{L}_2\left(k_{l, 1}/\mathbf{y}_{\mathrm{ch},l}\right).
\end{align}
We note that a solution to~\eqref{Equ:MLEst2}  can be attained using a brute-force search in a one-dimensional integer domain, which is much simpler than that in~\eqref{Equ:MLEst1}. Finally, by substituting the $\hat{k}_{l, 1}$ estimated from~\eqref{Equ:MLEst2} in~\eqref{Equ:h1}, we obtain the solution for $h_{l, 1}$ as
\begin{align}\label{Equ:XX}
   \hat{h}_{l,1}&=\begin{cases}
\frac{x^{\mathrm{H}}_p \boldsymbol{\Psi}^{\mathrm{H}}_{l, 1}\mathbf{y}_{\mathrm{ch},l}}{\left|\left| \boldsymbol{\Psi}_{l, 1}x_p\right|\right|^2}\Big|_{k_{l, 1}=\hat{k}_{l, 1}}, & \\
~~~~~~~~~~~~~~~\textrm{if}~\frac{x^{\mathrm{H}}_p \boldsymbol{\Psi}^{\mathrm{H}}_{l, 1}\mathbf{y}_{\mathrm{ch},l}}{\left|\left| \boldsymbol{\Psi}_{l, 1}x_p\right|\right|^2}\Big|_{k_{l, 1}=\hat{k}_{l, 1}}\geqslant \mathcal{T},\\
0, ~~~~~~~\textrm{otherwise},
\end{cases}
\end{align}
where $\mathcal{T}$ 
denotes the positive detection threshold used to avoid noise being falsely detected as propagation paths.

\section{Numerical Results}

In this section, we present numerical results to highlight the considerations of this work. 
The Extended Vehicular A model is adopted as the channel model while considering carrier frequency of $5~\mathrm{GHz}$, $\Delta f= 15~\mathrm{kHz}$, and user equipment speed of $500~\mathrm{km/h}$~\cite{2022_TWC_JH_ODDM}. Unless specified otherwise, the rest of the system parameters used for the numerical results are as follows: $N=M=128$, $S=8$, $T_{\mathrm{reg}}^{\mathrm{cp}}=\tau_{\mathrm{max}}=2.60 \mu \mathrm{s}$,
$T_{\mathrm{long}}^{\mathrm{cp}}=1.2\times T_{\mathrm{reg}}^{\mathrm{cp}}$,  $\hat{N}=20$~\cite{2019_TVT_YiHong_ChannelEstimationforOTFS,2018_TWC_Viterbo_OTFS_InterferenceCancellation,2017_WCNC_OTFS_Haddani,2022_TWC_JH_ODDM}.
Finally, we note that signal detection is performed in the DD domain using the state-of-the-art message passing algorithm~\cite{2018_TWC_Viterbo_OTFS_InterferenceCancellation}.

\begin{figure}[t]
\centering
\vspace{-1mm}
\includegraphics[width=0.85\columnwidth]{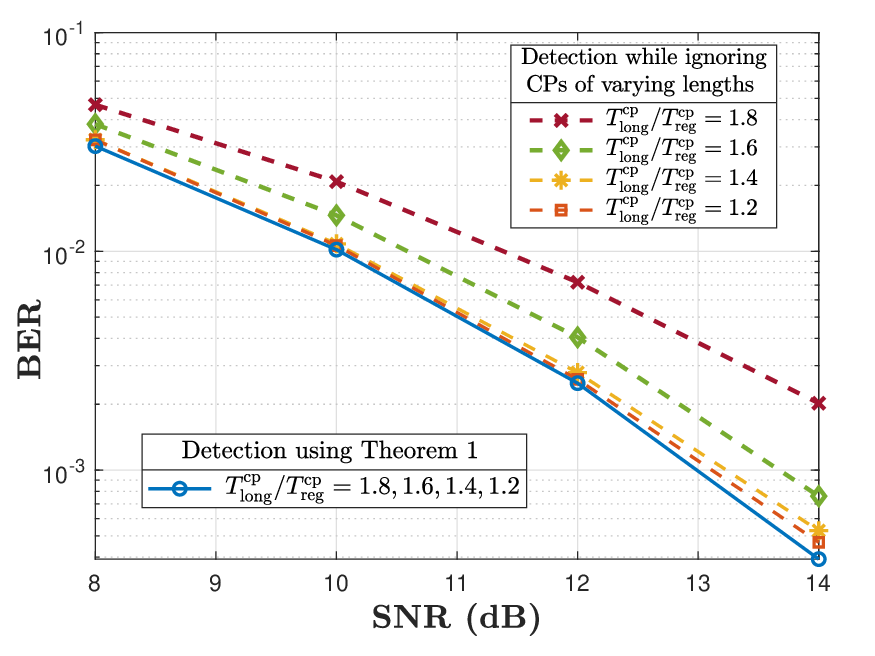}
\vspace{-4mm}
\caption{Illustration of the performance loss due to ignoring the unequal lengths of the CPs in the IOR derivation. }\label{Fig:NumFig2} \vspace{-3mm}
\end{figure}

Fig.~\ref{Fig:NumFig2} illustrates the impact of not ignoring the unequal lengths of the CPs when OTFS coexists with OFDM systems. 
To this end, we simulated OTFS in the considered coexisting system for different values of the ratio $T_{\mathrm{long}}^{\mathrm{cp}}/T_{\mathrm{reg}}^{\mathrm{cp}}$, and perform signal detection while using the channel matrix developed from the IOR derived in Theorem~1.\footnote{A single curve is plotted for different values of $T_{\mathrm{long}}^{\mathrm{cp}}/T_{\mathrm{reg}}^{\mathrm{cp}}$ for the scenario when signal detection is performed using the channel matrix developed from the IOR derived in Theorem 1. This is because for different values of $T_{\mathrm{long}}^{\mathrm{cp}}/T_{\mathrm{reg}}^{\mathrm{cp}}$, similar BER values are obtained when simulations are carried out for a sufficiently large number of  simulation trials.} 
For comparison, we perform signal detection on simulated OTFS signal while using the channel matrix derived in~\cite{2018_WCL_ModemStructureforOFDMBasedOTFS}, where the impact of unequal CP length is ignored.
First, it can be observed that for higher $T_{\mathrm{long}}^{\mathrm{cp}}/T_{\mathrm{reg}}^{\mathrm{cp}}$ values, e.g., $T_{\mathrm{long}}^{\mathrm{cp}}/T_{\mathrm{reg}}^{\mathrm{cp}}=1.6,1.8$, BER deterioration due to ignoring the impact of unequal lengths of the CPs  is very high. This shows that it may not be beneficial to ignore the impact of CPs of unequal lengths during detection when $T_{\mathrm{long}}^{\mathrm{cp}}/T_{\mathrm{reg}}^{\mathrm{cp}}$ is very high. However, we observe that the BER deterioration decreases when $T_{\mathrm{long}}^{\mathrm{cp}}/T_{\mathrm{reg}}^{\mathrm{cp}}$ decreases. As a result, when OTFS coexists with OFDM systems that use 5G NR numerologies which have low $T_{\mathrm{long}}^{\mathrm{cp}}/T_{\mathrm{reg}}^{\mathrm{cp}}$, it may be reasonable to perform signal detection while using the channel matrix in which the impact of unequal CP length is ignored. 
We clarify that this insight was made possible only because of our analysis on CP-OTFS with CPs of unequal lengths, thereby highlighting the significance of this work.

\begin{figure}[t]
\centering
\vspace{-1mm}
\includegraphics[width=0.85\columnwidth]{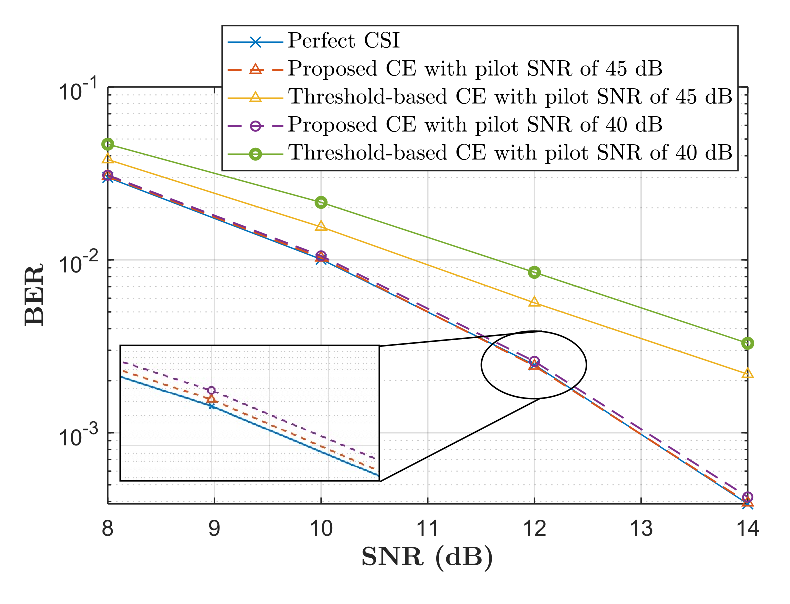}
\vspace{-4mm}
\caption{Illustration of BER versus SNR for different CE techniques.}\label{Fig:NumFig3} \vspace{-4mm}
\end{figure}

Next, to demonstrate the significance of the CE technique proposed in Section~\ref{Sec:CE}, Fig.~\ref{Fig:NumFig3} plots BER for the OTFS in the considered coexisting system versus SNR when the channel is estimated using the proposed CE technique. We consider $\mathcal{T}=3/\sqrt{\mathrm{SNR}_{\mathrm{p}}}$, where $\mathrm{SNR}_{\mathrm{p}}$ is the SNR of pilot symbols.
For comparison, we plot (i) the BER with perfect CSI and (ii)~the BER when the channel is estimated using the threshold-based CE technique proposed in~\cite{2019_TVT_YiHong_ChannelEstimationforOTFS}. 
We first observe that for all SNR values, the proposed CE technique outperforms the threshold-based CE technique. 
We also observe that the BER of the proposed CE technique approaches to that of perfect CSI. These observations show the significance of our proposed CE technique for OTFS in coexisting systems.
Moreover, we observe that although the BER of the threshold-based CE technique degrades significantly as the pilot power decreases, the deterioration in the BER for our proposed CE technique is marginal. This shows that it is possible to attain reasonable BER  for OTFS in coexisting systems without using high pilot power, thereby avoiding high peak-to-average power ratio during the implementation of OTFS in OFDM systems.

\section{Conclusion}

In this work, we investigated OTFS while focusing on its coexistence with current OFDM-based 4G/5G communication systems. We first derived the IOR of CP-OTFS while considering CPs of unequal lengths. We showed analytically that the inclusion of multiple CPs to the OTFS signal results in each propagation path being perceived as multiple taps with integer Doppler values, thereby leading to a less sparse effective sampled DD domain channel response. Thereafter, we proposed an embedded pilot-aided ML-based CE technique, which leverages the derived IOR. Using numerical results, we showed that it may not be reasonable to ignore the impact of CPs of unequal lengths during signal detection for OTFS in coexisting systems. We also showed that the proposed CE technique  outperforms the threshold-based CE technique.


\end{document}